# The systemic growth constants of climate change: From its origin in 1780 to its major post-WWII acceleration

Jessie Henshaw, HDS natural systems design science - sy@synapse9.com

## Abstract

The recent discovery of long term growth constants in the accumulation of atmospheric CO2, confirmed by two methods, enables analog methods for dating the beginning of climate change at ~1780 and projecting its near term future. Here we show that the preceding wavy variation in CO2 PPM abruptly shifts to exponential, moving symmetrically around a growth constant of 1.48 %/yr until WWII, and after a pause rises to hover about a higher constant growth constant from 1960 on of 2.0 %/yr. Such long term steady states of global environmental change suggest transitions between stable states of global self-organization. A method of analog curve fitting to project the current steady state of acceleration is tested, suggesting a 2040 earth temperature rise of 1.89 ºC above the IPCC baseline. A very brief following discussion of what a systemic growth constant for atmospheric CO2 implies and model strategies for responding to it.

Electronic supplementary material

Figures Slide set:

http://synapse9.com/_pub/GrowthConstantsOfClimate-figs.pdf  (Preliminary)

Preliminary Studies, Methods, Economic coupling:

http://synapse9.com/_pub/GrowthConstantsOfClimate-Supl.pdf  (Preliminary)

# First submission

# Systemic growth constants of climate change

The onset of climate change has been studied by Abram et al. ( 2016), finding it likely to have begun by ~1835.  To standardize temperature measures, the IPCC uses the 1850-1900 average temperature as a baseline, roughly agreeing with the wonderful study by Mann et al. (1998) showing the dramatic rise in temperatures breaking away from prior trends about 1900.  Neither study marks the beginning of climate change at the beginning of the greenhouse effect, used here to be able to look at it as a whole system.  The origins of this study and use of data curves to identify organizational states in natural systems originated with microclimate research in the 1970s (Henshaw 1978, 1979) and developed over the years into a general way of studying the organizational patterns of naturally occurring systems (Henshaw, 1985, 1999, 2008, 2010, 2011, 2015, 2018).  It was frustration with climate science not looking at the story of climate change from the beginning that led to this demonstration of the value of a whole system approach.  A good general reference to the subject of climate change is Hansen (2018)

This study uses pattern recognition to characterize the systems behind the data, starting with plotting the history of anthropogenic $CO_2$ back far enough (to 1500) to identify when the industrial use of fossil fuels started changing the global atmosphere (Fig 1).  Two different methods were used to identify the long period growth constant in the curve.  One was visually fitting growth curves to the data (manually adjusting variables for 'baseline,' 'rate' and 'exponent') (Fig 1), the other plotting locally averaged dy/Y growth rates of the data to be compared and adjusted to come into agreement (Fig 2).   Then an effort was made to understand the significant departures of the data from the idealized trends, resulting in ignoring some and adjusting the idealized trends to pass midway through local fluctuations, to see if the trends could represent a midline of local variation from a long term growth constant, presenting here those for which that method seemed successful.

The equations for the periods of steady-state growth are shown and then used for project the likely 2030, 40 and 50 temperatures (Fig 6) if the current steady state of economic fossil fuel use were to continue, showing somewhat a rate of change at the high end of the range shown in the recent IPCC projection to 2040 (1918).  Evidence of the global economy behaving as a whole and not a collection of separate economies.   The other half of the evidence that the world economy has states of steady self-organizing growth is simply the smoothness of the curve.  That shows that whatever combination of systems is reflected; they are all compensating for each other's behavior in a homeostatic way.  Presumably, that is through the economy working as it is supposed to, guided by constant economic market behavior for optimizing the use of every resource, the so-called "invisible hand" that moves all the parts to work together.

# Results

## The Origin of Global Warming

Climate change began with an economic "big bang" of using coal for steam locomotion and industries, clearly visible in the atmospheric CO2 PPM data emerging as an independent trend from background variation in about 1780 (Fig 1). A constant growth trend was visually fit by adjusting the constants till it seemed to be a centerline of all the small fluctuations in the first 160 years, finding a growth rate of 1.48 %/yr. At the start start there appears to be a jump of 3 PPM because the trend threads through the first fluctuation. You can see how the growth trend is so very different from lazy waves that preceeded 1780, and at the other end how at WWII the old trend vanished and there was a nominal 10 to 15 year unusual pause. That pause is then followed by an acceleration to a higher growth constant from about 1960 to the present, averaging about 2.0 %/yr when closely examined.

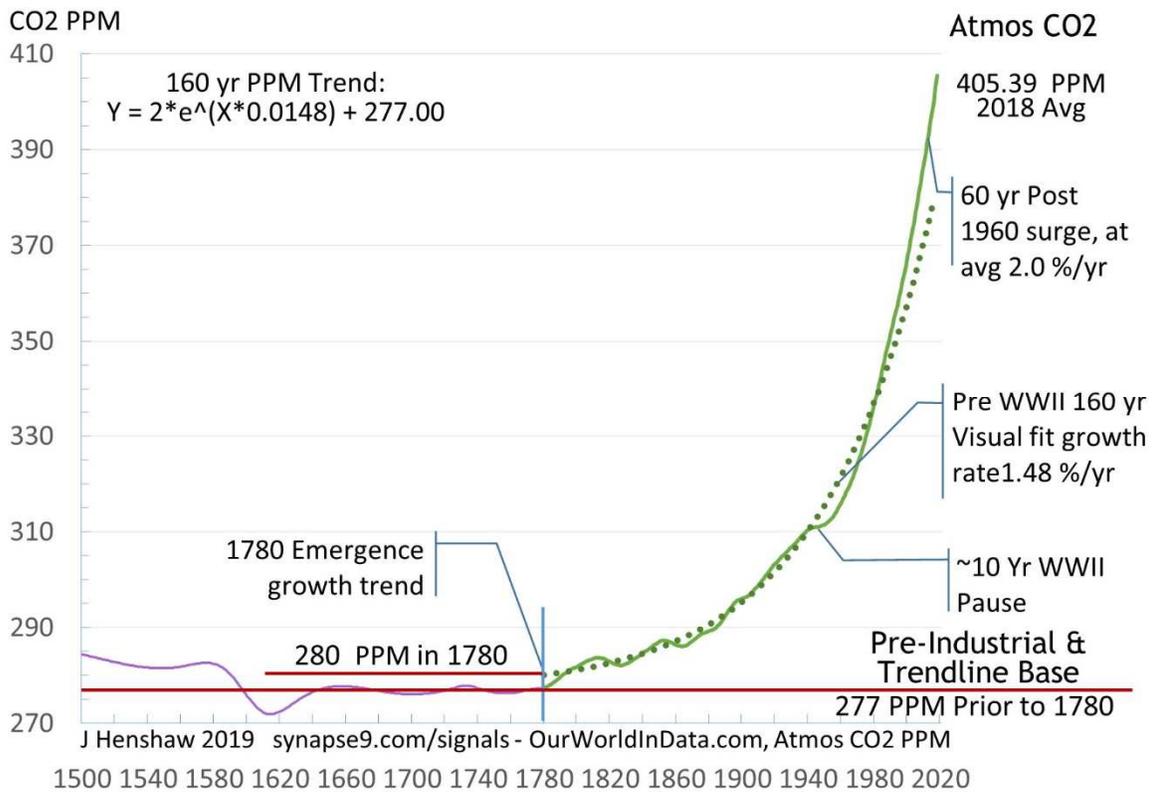

Fig 1.

Historical Atmospheric CO2 PPM from 1500; The wavy curve pre-WWII closely fits a constant 1.48%/yr growth constant and implies a 277 PPM pre-industrial baseline in 1780. The shape of the curve changes post-WWII to being unusually smooth and stabilizing at a 2.0 %/yr growth constant. What it represents historically seems likely to be the global systemization of economic growth post-WWII that we call globalization.

Visual fit 160 yr growth constant $\qquad Y = 2*e^{\wedge}(X*0.0148) + 277 \qquad$ (1)

# Systemic growth constants of climate change

That 160 yr growth constant (dotted line) is determined by visually fitting a growth curve to thread through the early fluctuations (adjusting the baseline, scale factor, and growth rate).   The industrial use of fossil fuels might well have had earlier formative periods as well; what we see at 1780 is the "big bang" of the systemization of growing fossil fuel use.  That the data curve seems to fluctuate symmetrically about a "growth constant" is also suggestive of homeostatic system organization, diverging and returning to that line of approximate symmetry again and again. That the underlying system was first interrupted and then successfully reorganized at a higher growth rate is particularly strong evidence of it being self-organizing and serially homeostatic.

It helps to tell any story from the beginning, setting the stage, and identifying the forces that will drive the narrative, even a data story like this.  The year 1780 was four years after the ratification of the US constitution, at a time of revolutionary economic change in the US and Europe.  Other important observations are that the overall smooth shape of the curve in Fig 1 represents a smoothly increasing climate forcing factor.  Fig 2 shows that the watts of greenhouse heating per square meter increase at a nominally linear rate in this period, making it possible to use the CO2 PPM curve as a proxy for earth temperature, if correctly adjusted to fit the history of the earth temperature data.   It's not a perfect proxy since at higher PPMs the relative heating tapers off a bit, and of course, because CO2 is only 50 to 60 % of the total GHG forcing of the climate (IPCC 2014).

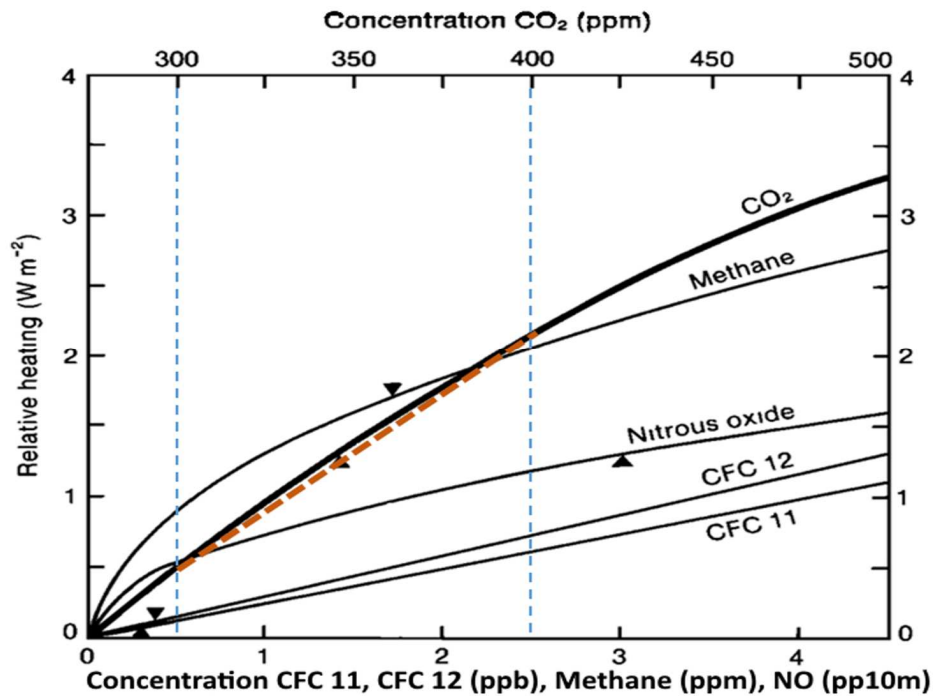

Fig 2.

Relative heating rates for CO2 PPM: Taken from Figure 6 in Mitchell (1989) "Greenhouse heating due to trace gases, showing [top scale] concentration of CO 2. […] The triangles denote 1985 concentrations".

The most significant irregularity in the PPM curve is the pause during and after WWII, as if world economic growth came nearly to a stop, and then even more rapidly accelerated, to stabilize at 2.0 %/yr.  It is a very rapid growth rate for a planetary system, *doubling* every 35 years now, not slowing down from the recent major improvements in efficiency but non-linearly accelerating.   It suggests that WWII allowed a general reorganization of the world economy for a faster growth rate of using fossil fuels, most likely the major scientific and institutional effort to accelerate growth we call "globalization."  All these details, the two growth constants, with each starting coincident with historical events, offer clear evidence of human "fingerprints" on the physical processes reflected in the data.   Because global warming is directly related to the greenhouse insulating effect of atmospheric $CO_2$ and other GHG gasses, global temperature rise would also have started as of 1780 too, and smoothly accelerated along with the $CO_2$ PPM.

The data also offers clear evidence that since the 1970s, when efforts to reduce impacts by improving energy efficiencies began in earnest, there has only been a rapid acceleration of the impacts that were supposed to decline.   That appears to be one main cause of our current unexpectedly rapid acceleration in climate change, that society fell prey to Jevons paradox (1885) that improving efficiency generally accelerates economic growth and so resource consumption and resulting pollution as well.

## Detailed CO2 growth rate movements

The Scripps $CO_2$ PPM data (Data Source #1) (Scripps 1958) (Macfarling Meur 2006) combines a sporadic ice core $CO_2$ data regularized with a spline from 1500 to 1958, followed by lightly smoothed annual average mountain-top air sample data from Mauna Loa and the Antarctic (Fig 3, lower curve).  Its dy/Y derivative growth rates (upper curve) show clearly the implied dramatic jumps in acceleration in the first 160 years replaced with a spline, displaying the sparse and highly variable ice core data as having dramatic positive and negative accelerations.  I was unable to discover any economic or natural phenomenon for such radical and falling of $CO_2$ in the atmosphere.   Looking at the sparse and highly irregular raw data in that period, I concluded that the large swings in growth rates were erroneous artifacts of splining the irregular ice core data. A separate study of the raw data (Preliminary 'Study A', Fig 7) helped confirm this.

In Fig 3 there are two exceptions, to the wild variation being erroneous, that the first acceleration is shown going off the scale, naturally infinite because of the abrupt start, and second the irregular ~30 yr slowing of $CO_2$ accumulation from WWI through the great depression and WWII.   The latter seems reinforced by 'Study A,' showing that the more frequent data in that period appears to move with the historical events, and reaching a real minimum at the end of WWII.  Post-1958, the dramatically better data shows a period of $CO_2$ PPM rising and leveling off at a new level with apparent damped oscillation.  That impression was also confirmed by 'Study A,' showing the amplitude of variation rising and falling with changes in scale, offering another sign of the stabilizing systemic organization.   I would have wanted to do a similar study of the historical $CO_2$ industrial emissions data but found evidence that' except from 1970 on, the available data CDIAC/NOAA (2015) is highly incomplete (see Prelim Studies B & C, Fig 8, 9, 10).

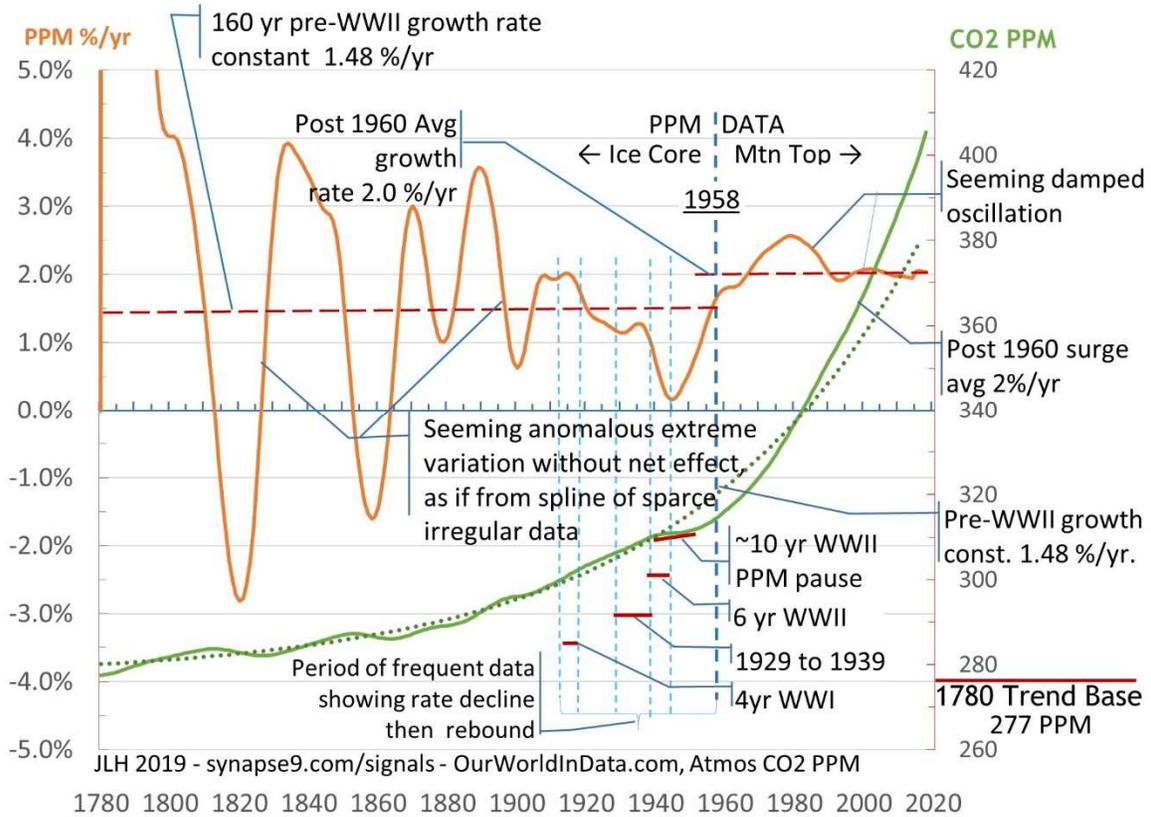

Fig 3.

The Scripps data on the Rt axis (lower curve Rt axis) is smoothed with a spline of the raw data to 1958 and a light running average after that. The dotted line is a visually fit 1.48 %/yr pre-WWII growth constant (Equn 1) with annual dy/Y growth rates (upper curve Lt axis) using a 5yr centered average(Equn 2). The dashed lines show the growth constants found that shift from before to after WWII.

Five point center-weighted smoothing    $f'(Y_n) = ( (Y_{n-2}+Y_{n-1}+Y_n)/3 - (Y_n+Y_{n+1}+Y_{n+2})/3 ) / Y_n$    (2)
applied to a time series of dy/y annual growth rates

## Fitting Climate Change & Atmospheric CO2

In Fig 4 the record of HadCRUT4 earth temperature data is shown adjusted to absolute °C units, using the estimated value of 14.6°C in 2017 as a scaling set point, assisted by Hawkins (2018). Below that is the same Atmospheric CO2 PPM data shown in Figs 1 & 3. In this case, the PPM curve is scaled to match the general shape of the °C earth temperature curve above to show the ways the cause and effect of climate change have similar shapes. The reason for shifting to abs °C units is to show the relationship between different baselines and eventually estimate the abs. Pre-Industrial temperature. The two curves differ quite a bit, too, but some of the important dynamics do appear to correspond. Both curves seem to display marked acceleration over time and to go from greater to smaller annual variation toward the end, as evidence of systemization. The big difference between the two curves is, of course, how irregular the annual earth temperature measurements are and how smooth the atmospheric CO2 curve is.

# Systemic growth constants of climate change

The physics is that CO2 is the main cause of the greenhouse effect, with a nominal linear relation, so theoretically the rate of increase in the earth's temperature should be linearly proportional to the concentration of CO2. So one can firmly conclude that the dramatic variation in the temperature record is due to something other than the greenhouse effect because the main forcing factor is increasing smoothly. I have not found studies that discuss irregular variations in annual temperature. I have assumed they were due to the interaction between atmospheric and ocean currents moving the incoming heat from the sun to different depths and elevations, such as El Ninos, not conforming with our placement of temperature measurements. A second difference between the curves is the presence of multi-decade "great waves" (1880, 1945) in earth temperature. You also see them throughout the long paleo history of earth temperature, Fig 5.

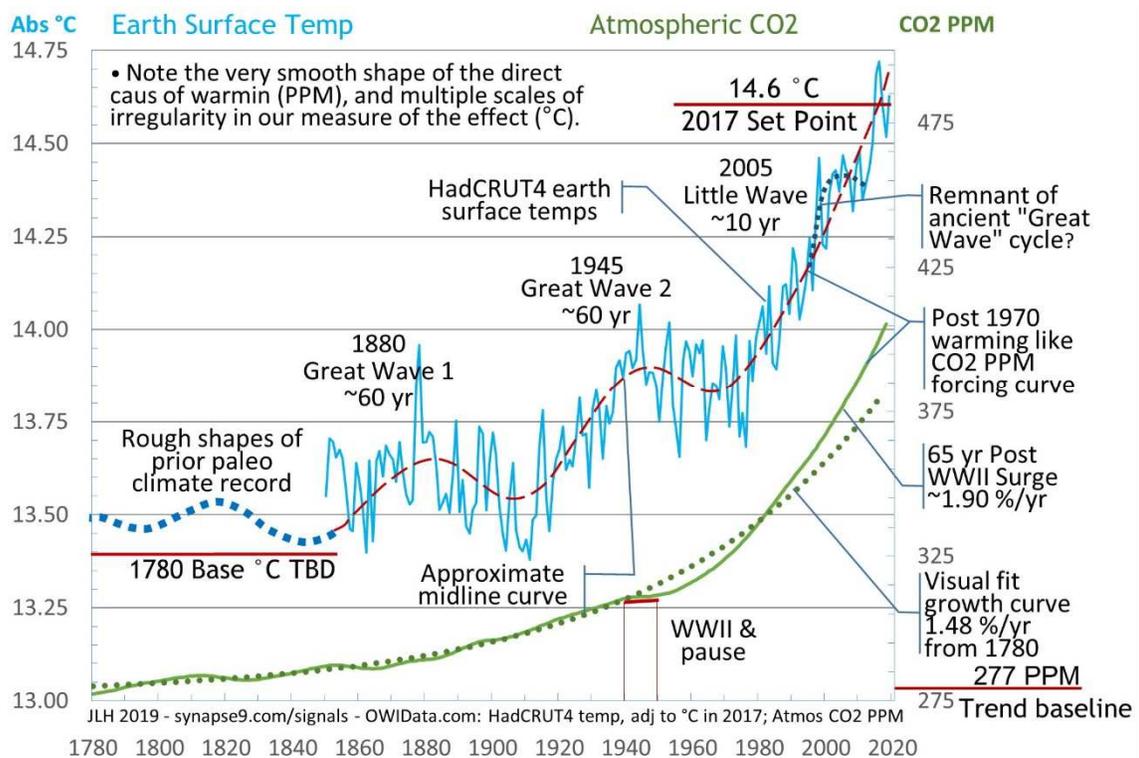

Fig 4.

Consideration of CO2 PPM and °C Coupling: The relation between the shapes of Global warming in °C and CO2 concentration in PPM might be consistent with their theoretically near-linear relation, depending on whether the departures can be explained and removed from the consideration.

To help tell the whole story, I have drawn a dashed midline through the rather erratic annual temperature fluctuations and extended that midline back to 1780 by imitating the general shapes of the great waves seen at that time in the NOAA paleoclimate record shown in Fig 5. I've also sketched in a "Little Wave" centered on 2005 to suggest it might be the remnant of a long great wave cycle, somehow weakened or interrupted by the dramatic acceleration of climate change after 1960. After thinking of a great many possible phenomena that could cause the long history of multi-decade waves in earth temperature, that could also now be interrupted by rapid warming; I came across only one that seemed plausible. For the

**Systemic growth constants of climate change**

purpose here it's only needed to make it plausible that the great waves are some climate cycle that stops working as warming intensifies. What I found plausible is that the great waves might represent multi-decade variation in equatorial stratosphere convection, creating long-standing radiative hotspots that irregularly ring the globe on both sides of the equator. Its speculation, but the idea is that the post-WWII, the CO2 PPM surge to growing at 2% a year, might be generating so much heat that the previous slowly pulsing high altitude convection cells would become permanent. These great waves seem not to be affected the concentration of CO2 at lower levels, though, only as the rise in PPM crosses a threshold and accelerates.

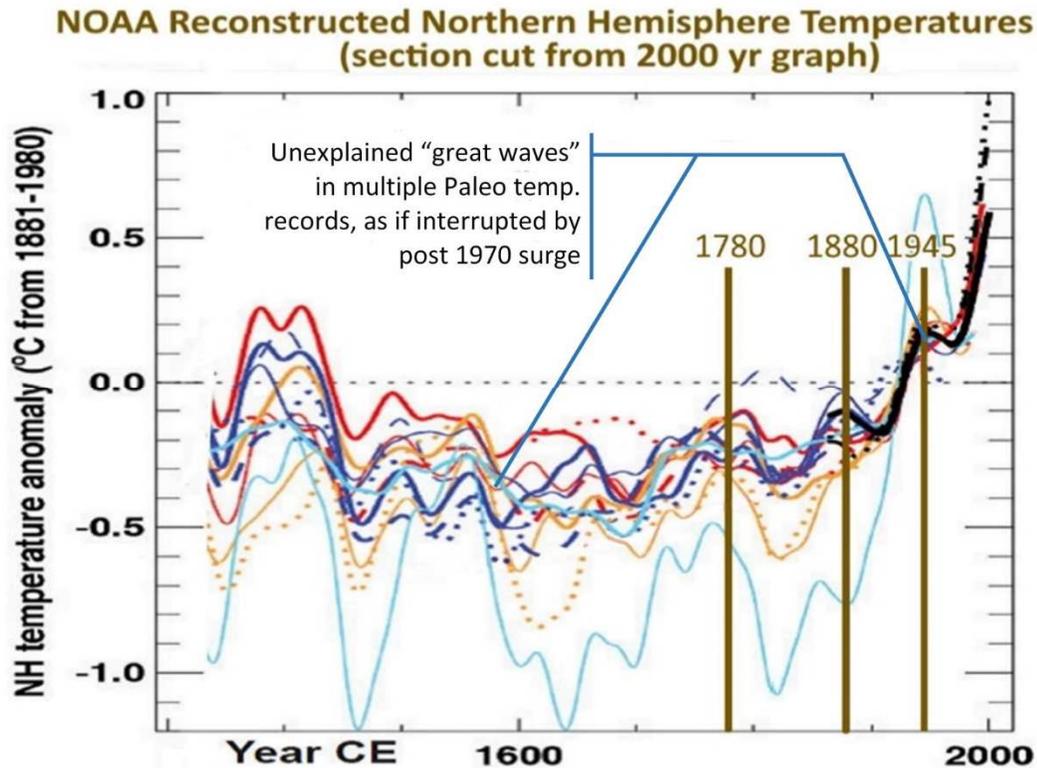

Fig 5.

A 900 yr portion of the NOAA (2007) 2000 yr Northern Hemisphere temperature record: A title and marks for 1780, 1880 and 1945 (brown) added, and an extraneous red line in the original was edited out. Note how the period of the exponential increase in Atmospheric CO2 from 1780 on is visible from following the great wave minima.

## Projecting Climate Change to 2040 and 2050

The IPCC has projected a 1.5°C rise in earth temperature by 2040 (IPCC 2018) but seems not to have factored in the possibility of a continuing 2 %/yr constant growth in CO2 PPM that the global economy appears to have been designed to produce. To check the IPCC's estimate in a way that should soon be testable, in Fig 6 I've scaled the baseline and scale CO2 PPM data curve with linear factors to achieve the visually best fit to the HadCRUT4 °C data. That has the effect of changing the PPM CO2 units into PPM°C temperature units data, needing to adjust the baseline to 13.27 °C to get the very best fit, as a first estimation of the true Pre-Industrial temperature in 1780. Then the PPM°C curve was projected to 2050 adding dy/Y

# Systemic growth constants of climate change

increments of 2 %/yr calculated from the 1780 baseline.  There would, of course, be a long list of possible other factors to consider.   However, because this method rests on the constants of the earth system as a whole for both the CO2 and temperature behaviors, all factors operating in the climate, such as cloud cover, and other GHGs influencing the earth system, are by default already accounted for.   Of course, one would like to find better data on the influence of the other GHGs, which might produce a combined global forcing factor curve, but that would be beyond the present scope in any case.

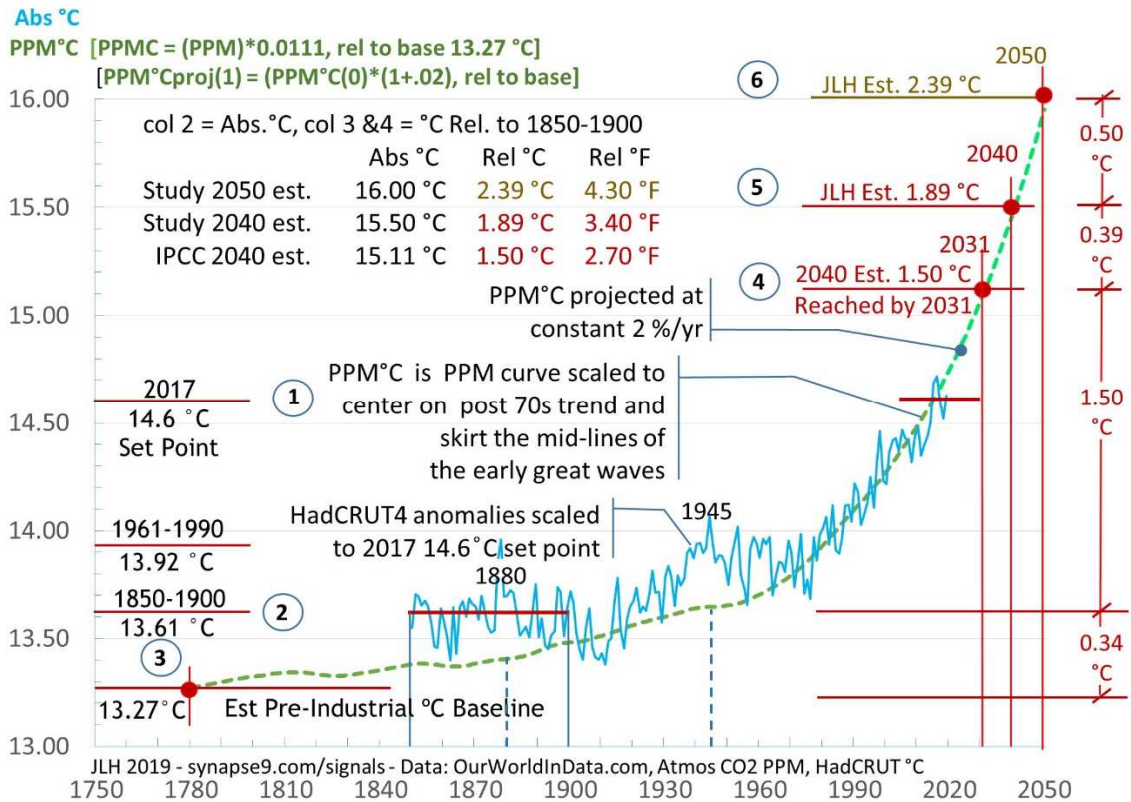

Fig 6.

2040 Projection - The history global temperature change; following 4 steps (1) Choosing an absolute temperature "set point" of 14.6 °C in 2017, (2) adjusting the HadCRUT4 anomalies to values of abs °C, (3) Visually shape-fitting a growth curve (r= .0143) from a 1780 base point origin (13.42 °C) to equally pass through the median of variations in the data, and (4) project the curve to 2040 to compare results with the IPCC's. Then study the other implications and adjust.

PPM converted to PPM°C   PPM°C  =  PPM*0.0111, relative to 13.27 °C base       (3)
        By adjusting the linear base and scale factors to best fit the HadCRUT4 data

PPM°C projection    PPMCproj(1) = (PPM°C(0)*(1+.02), relative to 13.27 °C base]    (4)

The big untested assumption that matters is how I fit the PPM°C trend line to the 1880 and 1945 "great waves" to follow their minima, not their midlines.  In part it was that only that assumption worked after trying lots of other possible ways to get the linear scaling of the PPM°C curve to closely fit the °C trend of the last ~50 years.  Here's where I use the speculation that the great waves might represent cycles of

J Henshaw                                                             9                                                              3-Nov-19

# Systemic growth constants of climate change

strengthening and weakening incursion of equatorial troposphere air masses into the stratosphere to speed up and slow down earth's radiative cooling. That what came to mind in looking for some process that for the great waves would "ride on top of." You can see in Fig 6 how the PPM°C curve is adjusted to pass through the minima on either side of the 1945 Great Wave, and approximately tangent to the midline of the annual fluctuations.

As you can see from Fig 6, instead of confirming the IPCC's estimate recent report for 1.5 °C by 2040, I found that 1.5 °C could come by 2031, that the 2040 temperature could rise to 2.15 °C, and by 2050, 2.40 °C. What makes the biggest difference is the established systemic growth rate of 2 ºC. Perhaps that was not being taken into account before would partly explain one of our more puzzling recent observations, that events have kept progressing faster than predicted (Dunlop & Spratt 2018).

One more strength of this method of curve fitting, putting together a "whole story" is that using the entire period from 1780 to the present gives the curve fitting a great many more fitting constraints, such as having to fit multiple scales of variation, greatly limiting the choices. I think that's what makes what I've called "visually fitting" a more than haphazard process, that you need to get all the scales of variation to fit, and have reasons for departing from that when you do, a rigorous process. That provides a kind of manual "regression" process guiding the adjustment of scaling factors by smaller and smaller amounts. Of course, having algorithms to suggest mathematical degrees of fit might help too but I don't think they'd not always do so well with fitting multiple scales of variation.

# Discussion

These kinds of observations are part of an exploratory method of identifying patterns and testing them on the way to build up to informative narratives and testable hypotheses. Having to do with development processes and their continuities the search goes back and forth between identifying continuities that seem to tell a story and then checking the departures, to see if they are "exceptions that help prove the rule" rather than "exceptions that invalidate the rule." That makes it unavoidable to carefully study each "bump on the curves" to extract the useful information about the organizational transformations reflected.

Why it seems unstudied I'm not sure, but lots of dynamic systems that maximize their acceleration seem to organize themselves around growth rate constants. As for $F = M*A$, the acceleration varies directly with the Force applied, making the maximum steady acceleration near the limits of the force the system can develop or take. The global system, tied to climate change that seems organized around growth constants that way is the world economy. Investors and businesses all seeking to maximize their rate of growing returns both stimulate maximum growth and punish less than that. The data showing it (Preliminary Study E and Fig, 11, 12, 13) suggests that "normal World GDP growth" of 3 ¼ %/yr has long been a "comfortable maximum acceleration" for investors, if less comfortable for others and the environment. That's hard to prove, more a matter of recognition, but at least it offers a reasonable hypothesis for what kind of self-organization produced the 1.48 %/yr and the 2.0 %/yr growth constants found in the study. My understanding of what happened in-between was that in the 1950s and 60s a world exhausted by war but technologically sophisticated, set about maximizing the world economic growth rate systematically, and succeeded. It included the creation of the modern networks of economic institutions, global economic cooperation driven by policy, science, business, technology, and finance that as a whole we now call "globalization." If the idea was to balance economic growth with the need to integrate environmental and human needs in that balance, now facing hard collisions with many of the planet's hard limits suggests we miscalculated.

So if a self-organized global system is driving climate change at ever faster rates, there may seem little one could do about it. It's just "the system." It's also a system that has clear patterns of organization of many kinds, some of which include the steering for how it chooses what directions it takes as it builds its future. Better-informed development decision making could change that functional steering of the economy toward maximizing sustainable profit, that doesn't bankrupt the system as a whole. Taking the common interests of the whole into account is unfamiliar, but that is what the growth system needs to recognize. For example, any economy needs to make a reliable profit, but due to fast approaching disruptions of life from climate change and a long list of other building global crises, we seem to be heading toward bankrupting the future instead. So to apply our ingenuity to this we would need to organize the great forces that built the world economy and its growth constants, to steer them and the world commons to a soft landing, using its economic institutions, international cooperation, science, business, technology, and finance, driven by self-interest and duty.

JLH

# Data Sources,

Note: The figures 7 to 14 are in the discussion are in the Supplementary Materials. An additional list of data sources is provided there.

1. Atmospheric CO2 PPM 1501-2015        Figs 1, 3, 6, 7, 10, 11, 12, 13, 14
   OurWorldInData.org:
   https://ourworldindata.org/co2-and-other-greenhouse-gas-emissions
   From Scripps source directly: (Scripps, 1958 to present)(Macfarling Meur 2006)
    http://scrippsco2.ucsd.edu/data/atmospheric_co2/icecore_merged_products
   "Atmospheric $CO_2$ record based on ice core data before 1958, and yearly averages of direct observations from Mauna Loa and the South Pole after and including 1958."

2. HadCRUT4 earth temperatures 1850-2017 –        Fit 5, 6
   Rosner - OurWorldInData.org: https://ourworldindata.org/co2-and-other-greenhouse-gas-emissions

# Systemic growth constants of climate change

# Systemic growth constants of climate change